\documentclass[usegraphicx]{mn2e}
\def\rc{R_{c}}
\def\rd{R_{d}}
\def\a0{a_0}
\def\ax{a_{{\rm max}}}

\begin{document}

\title[MOND halos]{MOND predictions of ``halo'' phenomenology in disc galaxies}
\author[M.Milgrom \& R.H. Sanders]
{Mordehai Milgrom$^1$ and Robert H. Sanders$^2$ \\$^1$ Department of
Condensed Matter Physics Weizmann Institute \\ $^2$ Kapteyn Astronomical
Institute, Groningen, NL}
\date{received: ; accepted: }
\maketitle
\begin{abstract}
We examine two corollaries of MOND pertaining to properties of the
equivalent dark-matter halo.
  MOND predicts for pure exponential discs a tight relation
 involving the halo and disc scale lengths and the mean
 acceleration in the disc, which we find to test favorably against
 the Verheijen sample of Ursa Major galaxies.
A correlation between halo and disc length scales is also apparent
when the ``maximum disc'' contribution is assumed, but we demonstrate
that this follows from the more general MOND prediction.
  The second MOND prediction involves the existence of a maximum
 halo acceleration, which also tests favorably against the Ursa Major
 sample for different assumptions on the disc contribution.

\end{abstract}


 \section{Introduction}
The quintessential test of MOND involves full analysis of rotation
curves of disc galaxies. MOND prescribes a simple formula whereby
the mass discrepancy in galaxies--as reflected in their measured
rotation curve--can be determined from the visible (baryonic) mass
distribution alone. For those who opt to view MOND as only a very
useful summary of dark-matter properties--a description that
``saves the phenomena" but is not underpinned by new physics--this
means that the distribution of baryons determines that of the dark
matter (DM).
 This sweeping prediction has been tested successfully in many galaxies
as reviewed, e.g., by Sanders \& McGaugh (2002), but in
it are buried a number of corollaries that deserve separate
discussion and testing.  These take the form of various predicted
properties of ``dark halos", correlations between visible matter
and ``dark matter" attributes, etc.. In MOND all these result from
one equation with one universal constant, as Newtonian dynamics
explain the different regularities in planetary motions.
 Without the umbrella of MOND these bylaws would
 appear as a set of many independent ``Kepler laws" of galactic
dynamics; see e.g., Milgrom (2002) for details.
\par
It is useful to consider such predictions on their own
 right because 1. they deal with more limited properties of
the mass discrepancy and hence are easier to take in; 2. detailing
them brings home the richness and variety of the predictions of
MOND; 3. they constitute intermediate challenges for dark matter
theories, which have, so far, anything but dealt with the tight
relation between baryonic and DM distributions encapsulated in MOND.
\par
An example of such a phenomenological bylaw brought to light by
MOND is the predicted correlation between rotational velocity and
{\it total} baryonic mass of a galaxy: the baryonic Tully-Fisher
relation. The innovation introduced by MOND is the inclusion of
the total visible mass comprising  that in gas as well as in
stars,  as stressed by Milgrom \& Braun (1988). It differs
from the standard TF relation, which involves only starlight, and
hence only stellar mass. This has been tested by McGaugh et al.
(2000) (and see also McGaugh 2004). Another
type of corollaries concerns the predicted shape of the DM
distribution in disc galaxies \cite{mil01}. Yet another is the
MOND prediction of the absence of a mass discrepancy in the body
of elliptical galaxies with high central surface densities
\cite{mil03}.
\par
 Here we discuss and
test two more such phenomenological laws predicted by MOND, which,
for pedagogical reasons, we
express as properties of an equivalent
 fictitious dark halo.  The first regularity,
examined here, in section 2, for the first time, concerns a
correlation between the disc and ``halo" scale lengths. The second
prediction, discussed in section 3, was noted by Brada \& Milgrom
(1999) but has not been tested before. It states that
there is an absolute maximum acceleration that a ``dark halo" can
produce.

\section{Correlation between halo and disc scale lengths}
Donato \& Salucci (2004) have recently found that in a sample
of disc galaxies, selected from several different sources on the
basis of the accuracy of the measured rotation curve, the core
radius of the best-fit pseudo-isothermal (PI) halo, $\rc$,
correlates well with the scale length of the stellar disc, $\rd$.
The correlation they find is of the form $\rc\approx 2\rd$. The
claimed tightness of the correlation is somewhat surprising as
halo core radii are notoriously difficult to pin down. Because of
the well known disc-halo degeneracy, when analyzing the inner
parts of rotation curves, the required halo core radii are usually
rather poorly determined without further assumptions on the
contribution of the disc \cite{ver97,veal03}. 
There are several possible ways of breaking this degeneracy:  for example,
one may assume a ``reasonable'' mass-to-light ratio for the stellar
disc, or that the disc makes its maximum possible contribution to
the rotation curve in the inner regions-- the so called ``maximum
disc hypothesis'' \cite {vasan}.  Donato \& Salucci take the
fitted halo parameters directly from the cited sources and used in
some cases the maximum-disc fits,
 while in other
cases the fits made with an assumed M/L for the disc,
 and for several objects the unconstrained minimum $\chi^2$
fits were used.
So, the role of these various assumptions on the
claimed correlation could not be readily assessed.

We have thus checked the correlation on another independent, more
homogeneous sample, the Ursa Major spirals observed by Verheijen
\cite{ver97}. Although the HI rotation curves are not all of the
highest quality, there are several advantages to using this
sample: e.g., all galaxies are roughly at the same distance (no
relative distance uncertainty), and there exist near-infrared (K'
band) surface photometry on all objects. The near infrared is less
susceptible to effects of dust obscuration and recent star
formation, thus eliminating an important source of uncertainly in
both the determination of $\rc$ and of $\rd$. In any event,
insistence on inclusion only of galaxies with very accurate
rotation curves is somewhat misplaced, as by far the most
prominent source of uncertainty in the present context is the
above noted disc-halo degeneracy.  We have, nonetheless, excluded
from the sample those galaxies designated by Verheijen as
``kinematically disturbed''.

As noted by Verheijen, when fitting pseudo-isothermal halos to the
rotation curves, $\chi^2$ minimization using all three parameters
indicates a very broad and noisy minimum. We therefore made
separate halo fits to the entire sample applying four different
assumptions about the disc contribution: maximum-disc, theoretical
M/L$_{K'}$ values obtained from the observed colors \cite{beleal},
an assumed constant M/L$_{K'}$ (=0.9), which is approximately what
the theoretical results give, and M/L$_{K'}$ values determined by
MOND one-parameter fits \cite{sv}. The number of objects (18-22)
included differs in the various cases because it was not possible
to achieve fits, using the PI halo, to all objects with, for
example, disc masses from fixed M/L.

\begin{figure}
\resizebox{\hsize}{!}{\includegraphics{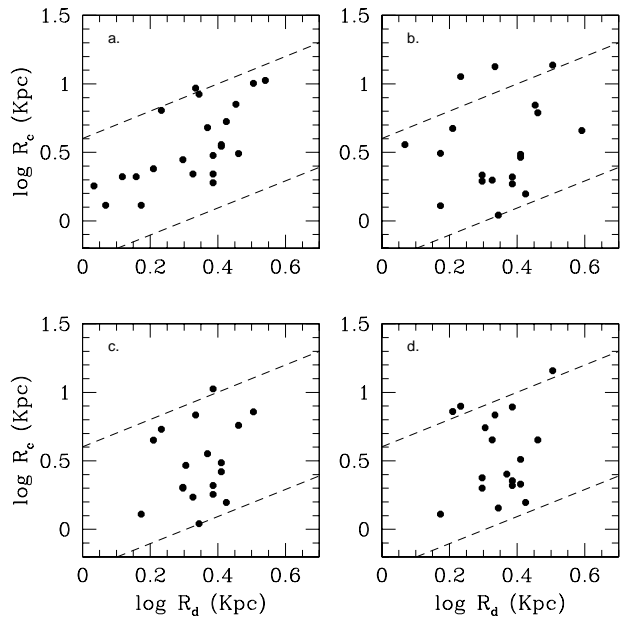}}
\caption[]{
Core radius of the best-fit isothermal sphere plotted against
the exponential scale length of the disc for Ursa Major spirals.
The different panels correspond to different assumptions on the
mass-to-light ratio of the stellar disc: a) maximum disc; b) $
M/L$ from population synthesis models \cite {beleal} and the
measured galaxy B-R colors \cite{ver97}; c) constant $M/L_K'~~(=0.9)$; 
d) best value from MOND fits.  The parallel dashed lines indicate 
$R_c/R_d$ = 0.5,4.0.}
\end{figure}

We plot in Fig.\ 1,
  the fitted halo core radius obtained
  under these four assumptions vs. the exponential-disc
scale length . It is evident that the only procedure giving an
apparent correlation is  the maximum-disc assumption (correlation
coefficient $\approx 0.7$).

The work of Donato and Salucci  has prompted us to check what MOND
says about such a correlation. Because in MOND the ``halo'' mass
distribution is determined by that of the visible matter, the
scale length of the ``halo'', in particular,
 depends on that of the baryonic galaxy.
But, there are other important factors that enter, which
 vary from galaxy to galaxy;
so, we do not expect a sharp correlation (as, e.g., is expected
 in the baryonic TF relation).
In the first place, the ``halo'' scale length depends also on the
mass distribution in the galaxy:
on the exact shape of the stellar
 mass distribution, on the relative contribution of gas and its
distribution, and on the presence and relative contribution of a
bulge. Even if we consider a sample of homologous baryonic
galaxies, such as pure exponential discs, there remains an
important disc parameter on which halo properties depend: the mean
acceleration in the galaxy in units of the MOND constant $\a0$.
\par
For the purpose of demonstration we shall, indeed, concentrate,
hereafter, on pure exponential discs. For these we use, as in
Milgrom (1983), the parameter $\xi\equiv v^2_{\infty}/\rd
\a0=(MG/\a0\rd^2)^{1/2}$ as the
 measures of the mean acceleration, where, here, $\rd$ is
 the exponential scale length, $v_{\infty}$ is the asymptotic rotational
 speed, and $M$ the total mass ($\xi$ is also a measure of the
mean surface density in units of $\a0/G$).
We then define the ``halo'' as the pseudo-isothermal
(PI) mass density
 distribution,
$$\rho(r)=\rho_0 \rc^2/(\rc^2+r^2),$$
that best completes the Newtonian rotation curve of the disc to
the MOND rotation curve. The ratio $Q\equiv\rc/\rd$ can then be
shown to
 depend on $\xi$ alone: $Q=Q(\xi)$.

\begin{figure}
\resizebox{\hsize}{!}{\includegraphics{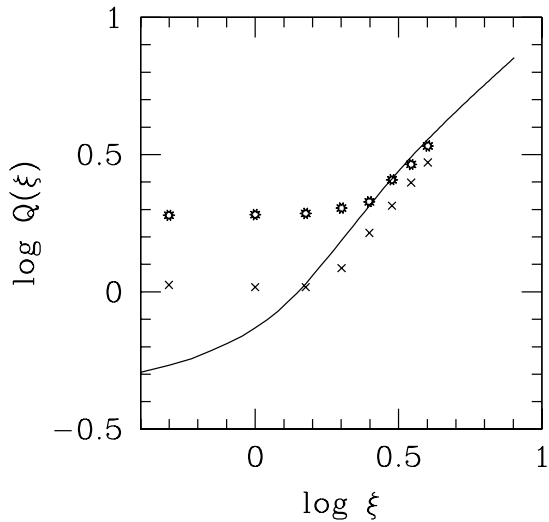}}
\caption[]{
The solid curve shows the MOND deduced ratio
 $Q\equiv\rc/\rd$ as a function of the mean
acceleration parameter $\xi$ for pure exponential discs.  The crosses
and open points show, respectively, the result for minimum $\chi^2$
and maximum disc fits to MOND rotation curves of exponential discs
 assuming Newtonian dynamics and a PI dark halo.}
\end{figure}

In the limit of  $\xi\ll 1$, i.e., deep in the MOND regime, the
MOND scaling laws dictate that $\rc$ becomes proportional to the
$\rd$. For exponential discs completed with PI halos we find
numerically in this limit $Q\approx 0.5$. As $\xi$ increases
beyond $\sim 1$ the inner parts of the disc become increasingly
Newtonian, the relative contribution of the halo there decreases,
and thus its core radius increases relative to $\rd$. In the limit
of high $\xi$, $Q(\xi)=q\xi$. This is true (with the same $q$) for
any galactic mass distribution (as long as it is bound) since in
this limit the ``halo" enters only at large radii where the galaxy
may be taken as a point mass; and, for a point mass $\rc\propto
r_t$, where $r_t\equiv (MG/\a0)^{1/2}=\xi\rd$ is the transition
radius.
 The constant $q$ is found numerically to be $q\approx0.9$.
A plot of $Q(\xi)$ for exponential discs is shown in Fig.\ 2
(assuming a MOND interpolating function of the form $\mu=
x/\sqrt{1+x^2}$).

\begin{figure}
\resizebox{\hsize}{!}{\includegraphics{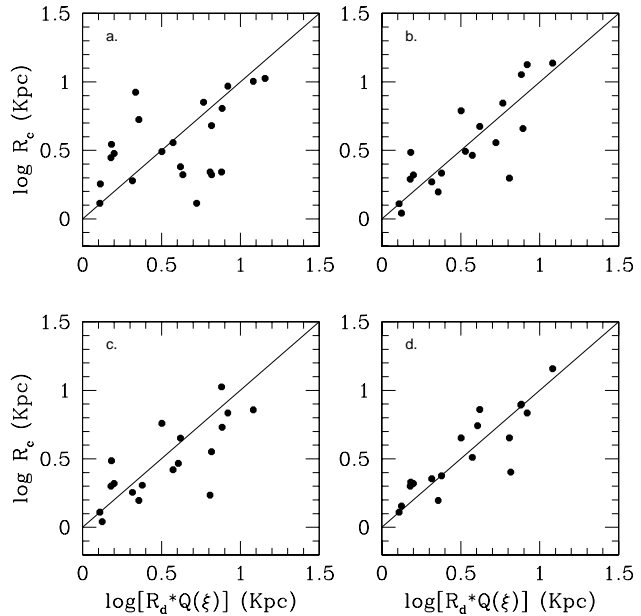}}
\caption[]{
The halo core radius, $\rc$, deduced by making different
 assumptions on the
 stellar $M/L$ values (as in the corresponding panels in Fig. 1)
  plotted against $\rd Q(\xi)$. The solid line is the MOND prediction.}
\end{figure}

\par
Galaxy discs have $\xi$ values that do not exceed $\sim 4$. This
is related to the generalized Freeman law whereby the distribution
of mean (or central) surface brightness of galactic discs is
cutoff from above at a value that corresponds, when converted to a
mean mass surface density, to $\sim \a0 G^{-1}$. At the other end,
there are low-surface brightness galaxies for which $\xi <1$.  So,
we expect $Q$ to range from 0.5 to $\sim 4$--as indeed is seen to
be the case in Fig. 1 for all but one assumptions on $M/L$. The
distribution of $Q$ for the maximum-disc assumption is narrower
(see below). So, rather than a tight correlation between $\rc$ and
$\rd$, we expect on the basis of MOND (for homologous discs)
 a ``fundamental surface'' in the
space $(\rc,\rd,\xi)$ given by $\rc=\rd Q(\xi)$.

Again, we test this prediction on the Ursa Major sample. Most of
the UMa galaxies are well represented by exponential
discs \cite{tv97}. An exception to this is NGC 3992-- a barred
spiral that does have a bulge, and whose total light distribution
is far from exponential. It was thus excluded from the sample.

We use the halo properties determined with the four different
choices of the stellar M/L values considered above. The results
are shown in Fig.\ 3. {\it Note that the solid line is not a fit
to this correlation, but the MOND prediction for the equivalent PI
halos}. The respective correlation coefficients are 0.48, 0.82,
0.77, 0.88. One might argue that the use of MOND disc masses
(Fig.\ 3d) is not an independent test of MOND predictions, but it
is evident that estimating the disc contribution from population
synthesis M/L's or even a constant M/L of 0.9 (in the K' band)
yields the expected correlation. These three assumptions on M/L
are based on some prior theoretical concept. We know from previous
analysis that MOND $M/L$ values are in good agreement with
theoretical values, and that both give a narrow distribution for
the $K$ band. So it is not surprising in light of this previous
knowledge, that all three give results consistent with each other.
The maximum disc assumption, while frequently applied, is rather
arbitrary, and is the one that gives a poor agreement with the
MOND prediction. This is expected in MOND: only galaxies with
high $\xi$ are expected to be correctly fit by a maximum disc.
 \par
 The question remains:  Why do Donato and Salucci obtain
 a good correlation between $\rd$ and $\rc$ without the scattering
 expected for a wide distribution of $\xi$, as do we for the
 Ursa Major sample when we apply the maximum disc assumption?
 We believe the reason rests in the following: The MOND ``halo'',
for $\xi<2$,
 is not typically well described by an isothermal sphere.
Thus, the best fit halo obtained by simultaneously fitting for the
disc $M/L$ and the halo parameters (the common practice) does not
give the same halo obtained from the best MOND fit.
 To simulate what is done in Newtonian fits we
should not take the MOND ``halo'' and fit it with an PI, but take
the full MOND rotation curve and perform the conventional
Newtonian fit on it. This we did for exponential discs with a wide
range of $\xi$ values.  We also applied a Newtonian maximum-disc
fit to this set. We calculated the $Q$ value obtained from these as
a function of $\xi$. The results are shown by the points in Fig.\
2 where the crosses show minimum $\chi^2$ fits and the open
points, maximum disc fits. It is evident that in both cases, Q
saturates at 1 and 2 respectively for $\xi<3$.  Thus the range of
Q predicted by MOND for maximum disc fits for example, is considerably
narrower than when using the correct disc contribution. The same
is true to a lesser extent for the best (three-parameter) fits.
This explains, in terms of MOND, the correlation found by DS: The
apparent relation $\rc \approx 2\rd$ is an artifact of the fitting
procedure which typically assumes maximum disc and therefore
artificially narrows the distribution of Q.

\section{Maximum halo acceleration}
Brada \& Milgrom (1999) have pointed out that the MOND
``halo'' acceleration in the plane of the galactic disc cannot
exceed a maximum value of order $\a0$. This results from the
relation
$$g_h(g)=g-g_N=g-g\mu(g/\a0),$$
between the  true (MOND) acceleration, $g$, and the halo
acceleration, $g_h$ (excess MOND over Newtonian), where $g_N$ is
the Newtonian acceleration, and $\mu$ is the interpolating
function of MOND. This expression is exact in MOND as modified
inertia \cite{mil94}, but is only approximate in modified gravity
MOND such as the formulation described by Bekenstein \& Milgrom
(1984). In any event, barring an anomalous behavior of
$\mu(x)$, $g_h$ cannot exceed a value of order $\a0$, call it
$\ax=\eta\a0$, where $\eta$ is constant of the theory (including
the exact choice of the
 function $\mu$). For the forms of $\mu$ used standardly
$\eta\approx 0.3-0.4$.

\begin{figure}
\resizebox{\hsize}{!}{\includegraphics{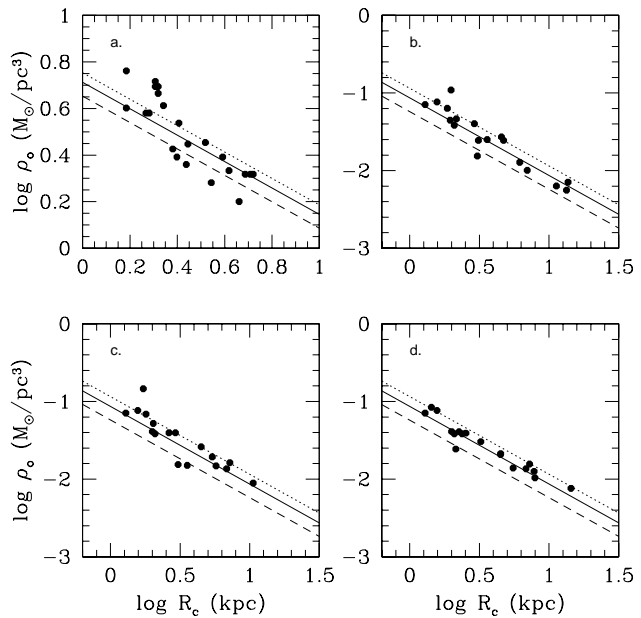}}
\caption[]{
The fitted halo central density ($\rho_o$) plotted against
the halo core radius, $\rc$ (log-log plot) where the disc
contribution is constrained via the four assumptions applied
previously (Figs.\ 1 and 3).  The parallel lines correspond to
maximum halo accelerations, the dashed line being $0.2\a0$,
the solid line $0.3\a0$, and the dotted line $0.4\a0$.} 
\end{figure}

 Such a maximum halo acceleration, if verified
observationally, would
 constitute a challenge for dark matter theories.
McGaugh (2004) has recently demonstrated, in fact, that
the above limit is inconsistent with the NFW halo parameters
deduced from CDM N-body simulations, which give too high halo
accelerations for high-surface-density galaxies.

We test the maximum-halo-acceleration on best fit PI halos. For
such halos the acceleration goes to zero at small and large radii,
and reaches a maximum at $\approx 1.5\rc$, where its value is
$$ a_{max}\approx 2.9 G\rho_0\rc.$$

Again we consider the PI fits to UMa spirals with the four
assumptions on the stellar contribution as in Figs. 1 and 3. A
plot of the resulting central halo density, $\rho_o$ vs. the halo
core radius, $\rc$ is shown in Fig. 4 along with lines of constant
maximum halo acceleration given by the expression above. The $\pm
1\sigma$ errors on the fitted core radius and density are
typically 30\% and  50\% respectively, but these errors are
correlated: the error ellipses are extended roughly parallel to
the lines of constant $\rho_c{R_c}^{1.5}$.
 Nonetheless, it is evident that the equivalent PI
halos do seem to exhibit a characteristic maximum acceleration
($g_h<0.4 \a0$) as predicted by MOND.

\section {Conclusions}

The simple MOND formula makes many definite predictions of
regularities in the properties of the equivalent, Newtonian
pseudo-isothermal halos, two of which we have considered here. The
first is that, when the galaxies are pure discs, and the surface
density distribution is well described by a simple exponential,
one expects a tight correlation involving the disc length scale,
the core radius, and the mean-acceleration parameter. The core
radius is not expected to be well correlated to the disc
radius for a sample
that involves a spread in the mean acceleration, unless one makes an
assumption, such as maximum disc, which artificially restricts the
distribution of $\rc/\rd$.  From Figs. 1 and 3
we see that this prediction appears to be borne out for the UMa
sample.

The second prediction is that there exists a maximum possible halo
acceleration, which depends somewhat upon the form of the MOND
interpolating function $\mu$ but is typically less than $\a0$.  In
Fig.\ 4 it is evident that fitted halo core radii and central
densities are indeed such that the halo acceleration is always
less than $\sim 0.4\a0$, as predicted.

We are grateful to Stacy McGaugh for helpful comments on the
manuscript.

\clearpage
\end{document}